\documentclass[journal=nalefd,manuscript=article]{achemso}

\usepackage[T1]{fontenc}       
\usepackage{graphicx,color}
\usepackage{amsmath}
\usepackage{amsfonts}
\usepackage{amssymb}

\author{Minkyung Jung}
\email{minkyung.jung@dgist.ac.kr/Tel: +8253785 3501, Fax: +82 53 785 3439}
\affiliation[DGIST]
{DGIST Research Institute, DGIST, 333 TechnoJungang, Hyeongpung, Daegu 42988, Korea}

\author{Kenji Yoshida}
\affiliation[The University of Tokyo]
{Center for Photonics Electronics Convergence, IIS, University of Tokyo, 4-6-1 	Komaba, Meguro-ku, Tokyo 153-8505, Japan}

\author{Kidong Park}
\affiliation[Korea University]
{Department of Chemistry, Korea University, Sejong 339-700, Korea}

\author{Xiao-Xiao Zhang}
\affiliation[The University of Tokyo]
{Department of Applied Physics, The University of Tokyo, 7-3-1 Hongo, Bunkyo-ku, Tokyo 113-8656, Japan}

\author{Can Yesilyurt}
\affiliation[National University of Singapore]
{Electrical and Computer Engineering, National University of Singapore, Singapore, Republic of Singapore 117576}

\author{Zhuo Bin Siu}
\affiliation[National University of Singapore]
{Electrical and Computer Engineering, National University of Singapore, Singapore, Republic of Singapore 117576}

\author{Mansoor B. A. Jalil}
\affiliation[National University of Singapore]
{Electrical and Computer Engineering, National University of Singapore, Singapore, Republic of Singapore 117576}

\author{Jinwan Park}
\affiliation[DGIST]
{Department of Emerging Materials Science, DGIST, 333 TechnoJungang, Hyeongpung, Daegu 42988, Korea}

\author{Jeunghee Park}
\affiliation[Korea University]
{Department of Chemistry, Korea University, Sejong 339-700, Korea}
	
\author{Naoto Nagaosa}
\affiliation[The University of Tokyo]
{Department of Applied Physics, The University of Tokyo, 7-3-1 Hongo, Bunkyo-ku, Tokyo 113-8656, Japan}
\alsoaffiliation[RIKEN]
{RIKEN Center for Emergent Matter Science (CEMS), 2-1 Hirosawa, Wako, Saitama 351-0198, Japan}

\author{Jungpil Seo}
\email{jseo@dgist.ac.kr}
\affiliation[DGIST]
{Department of Emerging Materials Science, DGIST, 333 TechnoJungang, Hyeongpung, Daegu 42988, Korea}
	
\author{Kazuhiko Hirakawa}
\email{hirakawa@iis.u-tokyo.ac.jp}
\affiliation[The University of Tokyo]
{Center for Photonics Electronics Convergence, IIS, University of Tokyo, 4-6-1 	Komaba, Meguro-ku, Tokyo 153-8505, Japan}
\alsoaffiliation[The University of Tokyo]
{Institute for Nano Quantum Information Electronics, University of Tokyo, 4-6-1 Komaba, Meguro-ku, Tokyo 153-8505, Japan}

\title{Quantum dots formed in three-dimensional Dirac semimetal Cd$_3$As$_2$ nanowires}


\begin{document}

\begin{abstract}
	
We demonstrate quantum dot (QD) formation in three-dimensional Dirac semi\-metal Cd$_{3}$As$_{2}$ nanowires using two electrostatically tuned p$-$n junctions with a gate and magnetic fields. The linear conductance measured as a function of gate voltage under high magnetic fields is strongly suppressed at the Dirac point close to zero conductance, showing strong conductance oscillations. Remarkably, in this regime, the Cd$_{3}$As$_{2}$ nanowire device exhibits Coulomb diamond features, indicating that a clean single QD forms in the Dirac semimetal nanowire. Our results show that a p$-$type QD can be formed between two n$-$type leads underneath metal contacts in the nanowire by applying gate voltages under strong magnetic fields. Analysis of the quantum confinement in the gapless band structure confirms that p$-$n junctions formed between the p$-$type QD and two neighboring n$-$type leads under high magnetic fields behave as resistive tunnel barriers due to cyclotron motion, resulting in the suppression of Klein tunneling. The p$-$type QD with magnetic field-induced confinement shows a single hole filling. Our results will open up a route to quantum devices such as QDs or quantum point contacts based on Dirac and Weyl semimetals.\\





\end{abstract}

\newpage

Three-dimensional (3D) Dirac semimetals ($e.g.,$ Cd$_{3}$As$_{2}$ and Na$_{3}$Bi) have attracted considerable attention owing to their exotic properties, both predicted in theory\cite{Young_PRL2012,Wang_PRB2012,Wang_PRB2013} and recently demonstrated experimentally.\cite{Liu_Science2014,Liu_Nat_mat2014,Neupane_Nat_comm2014,Borisenko_PRL2014, Xu_Science2015} They are a 3D analogue of graphene, but the Dirac points of these semimetals are not gapped by the spin-orbit interaction and the crossing of the linear energy band dispersion relations along all three momentum directions is protected by  their crystal symmetry.\cite{Young_PRL2012} This unique energy band structure makes Dirac semimetals not only a source of intriguing physical phenomena ($e.g.$, giant magnetoresistance (MR),\cite{Liang_Nat_mat2015,Feng_PRB2015,Narayanan_PRL2015,Shekhar_nat_phy2015} ultrahigh carrier mobility,\cite{Liang_Nat_mat2015} Landau quantization,\cite{Jeon_Nat_mat2014,Cao_Nat_comm2015} and the Shubnikov-de Haas effect\cite{Feng_PRB2015,Narayanan_PRL2015,He_PRL2014} in 3D bulk Dirac semimetals), but also a new potential platform for quantum devices.   

Recently, 3D Dirac semimetal Cd$_{3}$As$_{2}$ nanowires have been successfully grown by chemical vapor deposition (CVD).\cite{Li_Nat_comm2016,Li_Nat_comm2015,Wang_Nat_comm2015,Wang_PRB2016} One-dimensional nanowires with large surface-to-volume ratios are an excellent platform to explore novel physics and device applications that are difficult to achieve with bulk materials.\cite{Lieber_MRS2007} In Dirac semimetal Cd$_{3}$As$_{2}$ nanowires Aharonov-Bohm (AB) oscillations, which are an indicator of probing the surface states of the Dirac semimetal, have been observed.\cite{Wang_Nat_comm2015} A phase shift in AB oscillations is attributed to the splitting of the Dirac node into Weyl nodes owing to magnetic field-induced time-reversal symmetry breaking. In addition, other unique features such as a large negative MR and universal conductance fluctuations have been observed.\cite{Li_Nat_comm2016,Li_Nat_comm2015,Wang_PRB2016} 


However quantum dot (QD) formation has not yet been demonstrated in 3D Dirac materials because the gapless band structure of the Dirac semimetals prevents quantum confinement which is essential for QD formation. Nevertheless, quantum confinement can be accomplished through lithographically patterned structures, as achieved in graphene devices.\cite{Guttinger_Review2012} This system, however, is inevitably contaminated or the edges of the defined areas are too rough, producing random potential fluctuations that make devices unstable. An alternative approach for confining electrons in the Dirac materials is to use electrostatic potentials ($e.g.$, p$-$n junctions). However, this has been difficult to achieve because massless Dirac fermions can penetrate p$-$n junctions with high transmission probability at non-oblique incidence. This is referred to as Klein tunneling.\cite{Katsnelson_Nat_phy2006,Young_Nat_phy2006,Cheianov_PRL2006}

\begin{figure}[t]
	\centering
	\includegraphics[width=0.8\textwidth]{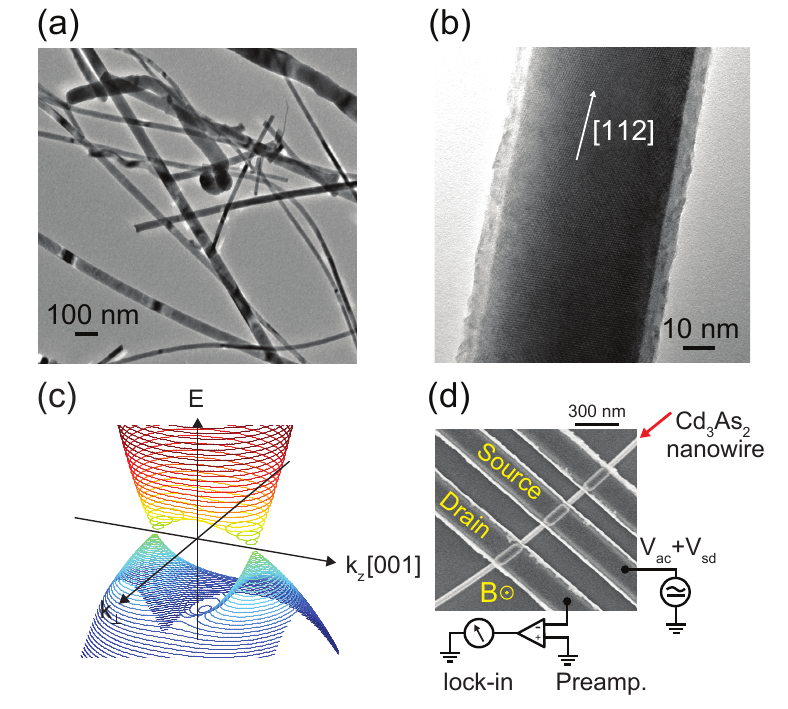}
	\caption{(a) TEM image of 3D Dirac semimetal Cd$_3$As$_2$ nanowires. (b) High-resolution TEM image of a nanowire indicates $[112]$ growth direction. The surface of nanowires is covered by several amorphous layers. (c) Schematic of the energy dispersion $E(k)$ near the Dirac nodes in Cd$_3$As$_2$. (d) SEM image and measurement setup of the Cd$_3$As$_2$ nanowire device with multiple source and drain contacts. The channel length and diameter of the nanowire are $\sim250$ nm and $\sim70$ nm, respectively. The magnetic field is applied perpendicular to the nanowire.}
\end{figure}

Here, we report on QDs formed in 3D Dirac semimetal Cd$_{3}$As$_{2}$ nanowires with magnetic confinement. The linear conductance measured as a function of gate voltage with increasing magnetic field is strongly suppressed at the Dirac points that are close to zero. Under high magnetic fields, the device exhibits a single QD behavior, demonstrating a clean charge stability diagram. Unlike normal semiconductor nanowires, bipolar transport properties such as p or n$-$type can be easily probed in the Dirac semimetals by controlling gate voltage due to absence of energy band gap.

Cd$_3$As$_2$ nanowires are synthesized by the vapor transport method, and they have large surface-to-volume ratios. Cd$_3$As$_2$ (99 \%, Alfa Aesar) powder is placed in ceramic boats, which are loaded inside a quartz tube reactor. A silicon substrate coated with 1 mM BiI$_3$ (99.999 \%, Sigma-Aldrich Corp.) in ethanol solution to form the Bi catalytic nanoparticles is positioned 8 cm away from the powder source. Argon gas is continuously supplied at a rate of 200 sccm under ambient pressure during growth. The temperature of the powder sources is set to $450$\,$^{\circ}$C. The substrate is maintained at $T$ $\sim$ 350\,$^{\circ}$C to synthesize the nanowires. Transmission electron microscopy (TEM) images are shown in Figure~1a and~1b. The TEM image in Figure~1b shows that nanowires grow preferentially along the [112] direction, which is the axial direction (Supporting Information, Figure S1). The surface of nanowires is covered by several amorphous layers in the ambient condition. Figure~1c shows a schematic of the energy dispersion $E(k)$ near the Dirac points in the Cd$_3$As$_2$ semimetal. It is known that in a Dirac semimetal the conduction and valence bands contact only at the Dirac points in the Brillouin zone and they disperse linearly in all three directions,\cite{Young_PRL2012,Neupane_Nat_comm2014} resulting in large Fermi velocity and very high electron mobility of the 3D carriers.\cite{Jeon_Nat_mat2014,Liang_Nat_mat2015}

A scanning electron microscopy (SEM) image of a nanowire device and the measurement setup are shown in Figure~1d. The Cd$_3$As$_2$ nanowires are transfered on a highly doped p$^{++}$ silicon substrate covered by 300 nm-thick thermally grown SiO$_2$ using a micro-manipulation technique. The p$^{++}$ silicon is used as a global backgate. The typical diameter of nanowires selected for device fabrication is $\sim60-100$ nm, as measured by SEM. Ti/Au (5/100 nm) contacts separated by 250 nm are fabricated after an Ar plasma etch to remove the native oxide at the surface of the nanowire. The device is measured at a temperature of $\sim 300$ mK. The magnetic field ($B$) is applied perpendicular to the nanowire, that is, $B$$\perp$$E$ (electric field).

\begin{figure}[t]
	\centering
	\includegraphics[width=0.7\textwidth]{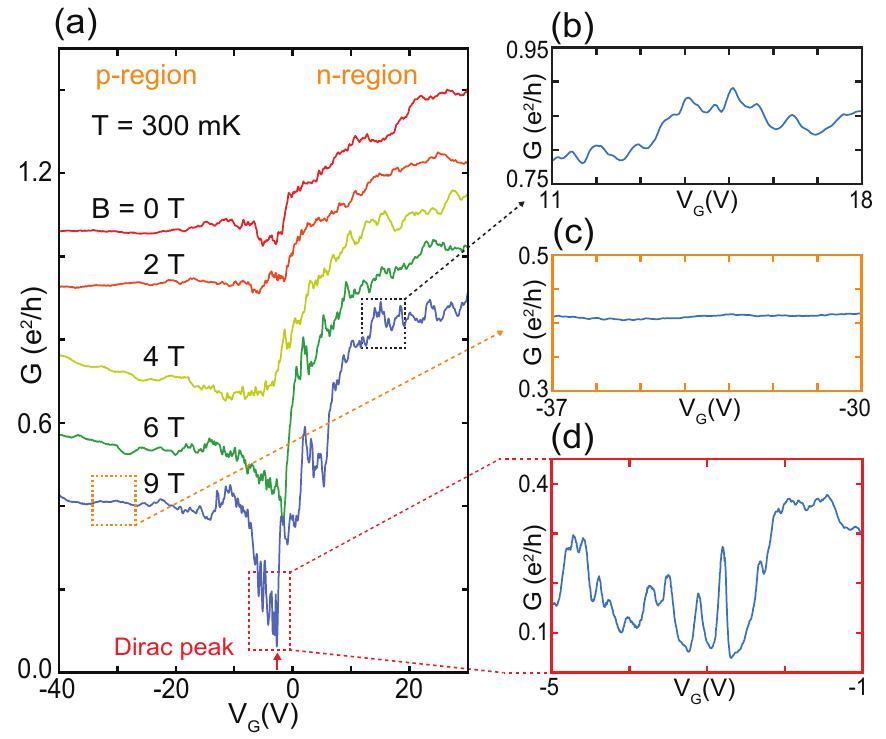}
	\caption{(a) Differential conductance $G$ as a function of backgate voltage $V_{G}$ at a source-drain voltage $V_{SD}$ = 0 V for different magnetic fields. The conductance decreases with an increasing magnetic field $B$. The device shows ambipolar behavior, showing the Dirac peak at $V_G =$ $\sim-$2.5 V, as marked with a red arrow. Magnified conductance traces at (b) positive high gate voltages, and (c) negative high gate voltages for $B = $ 9 T. Aperiodic and reproducible oscillations superimposed on the high conductance in the n-region can be attributed to universal conductance fluctuations (UCFs) (panel b), while no conductance oscillations are observed in the p-region, as shown in panel (c), a magnified area of the orange dashed box in (a). (d) Differential conductance at $B = $ 9 T measured at the Dirac peak, as indicated by the dashed box in (a). Strong conductance oscillations that are distinguishable from UCFs are observed.}
\end{figure}


Figure~2a shows the differential conductance $G$ versus backgate voltage $V_G$ under various magnetic fields ranging from 0 to 9 T. By sweeping the gate voltage, the device exhibits ambipolar behavior, indicating the presence of 3D Dirac points of the Cd$_3$As$_2$ nanowire. The minimum conductance is identified at $V_G$ $\sim-2.5$ V (marked with a red arrow in Figure~2a) as the location of the Dirac points for $B = $ 9 T. At $B = $ 0 T, we observe that the gate dependence of the conductance is relatively weak and it becomes significant under high magnetic fields. This is because the conductance near the Dirac point is markedly reduced as the magnetic field increases. This tendency is seen more clearly in the MR plot (Supporting Information, Figure~S2b). The MR plot shows a strong magnetic field dependence as it increases exponentially up to 1700 \% at $B =$ 9 T. In our experiment, we have not observed particular evidence of the surface states. Their contribution to conductance compared to the bulk could be small when considering the density of states of the surface states and bulk.



It should be noted that small aperiodic conductance oscillations are superimposed on the background conductances in the n and p$-$regimes as shown in Figure~2. Figure~2b shows a magnified conductance trace taken at positive high gate voltages. The oscillations are very reproducible with gate voltages and magnetic fields, and they do not show hysteresis in the conductance. Their origin can be attributed to universal conductance fluctuations (UCFs), a signature of phase coherent transport for electrons between the source and drain electrodes.\cite{Wang_PRB2016} Notably, the UCFs are significantly suppressed in the deep p$-$regime ($V_G < \sim -$20 V) as a magnified trace is shown in Figure~2c (dashed orange box). Two possibilities are presented to account for this. First, the coherence length for holes could be much shorter than that of electrons, which is consistent with the fact that electron mobility ($\mu_e = 1300$ cm$^2$/Vs) is approximately six times higher than that of the hole ($\mu_h = 220$ cm$^2$/Vs) in this devices.\cite{Wang_PRB2016} Second, as will be discussed below (Figure 4), since the device forms p$-$n junctions between the nanowire and the metal electrodes by applying negative gate voltages, quantum interference originating from coherent propagation of holes is suppressed at two p$-$n junctions. With increasing temperature, the oscillation amplitude is suppressed (Supporting Information, Figure S3). Interestingly, we observe strong oscillations with much larger amplitudes near the Dirac peak under high magnetic fields (Figure~2d). In contrast to UCFs, the oscillations are only pronounced at high magnetic fields ($B >$ 7 T), indicating that the origin should be different from UCFs.

\begin{figure}[t]
	\centering
	\includegraphics[width=0.8\textwidth]{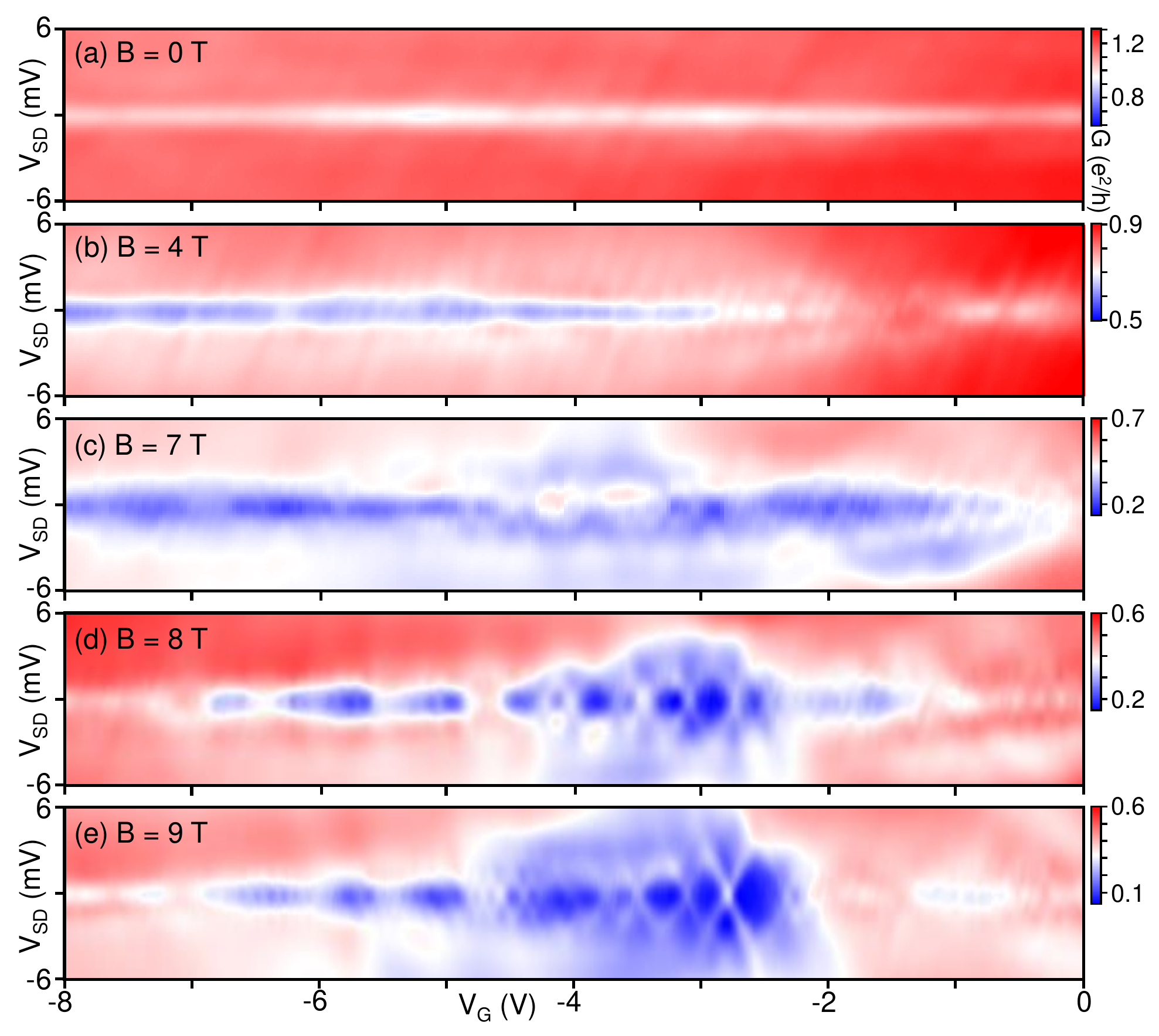}
	\caption{Differential conductance $G$ in units of e$^{2}$/h measurements as a function of $V_G$ and $V_{SD}$ at (a) $B =$ 0 T, (b) $B =$ 4 T, (c) $B =$ 7 T, (d) $B =$ 8 T, and (e) $B =$ 9 T. The entire conductance is reduced as the magnetic field $B$ increases. Surprisingly, from approximately $B =$ 8 T, the conductance maps show clear diamond shaped areas between $V_G = -$2 V and $-$5 V, which can be attributed to Coulomb diamonds. This indicates that a single QD is formed between the source-drain contacts, and confirms that the strong conductance oscillation pattern observed in Figure~2d is a Coulomb oscillation.}
\end{figure}

The two-dimensional conductance maps as a function of magnetic field provide insight into the strong conductance oscillations at around the Dirac peak, as seen in Figure~2d. The differential conductance maps measured as a function of $V_{G}$ and $V_{SD}$ at $B =$ 0, 4, 7, 8 and 9 T are presented in Figure~3a$-$e, respectively. For $B =$ 0 T, the device exhibits very weak aperiodic oscillations fluctuating from 0.9 to 1.3 e$^{2}$/h, which are attributed to UCFs as mentioned above. However, at high magnetic fields ($B \gtrsim$ 8 T), the conductance maps show clear diamond shaped areas at approximately zero bias, which can be attributed to the Coulomb stability diagrams. As shown in Figure~3e, the conductance lines along the diamond edges are straight, and Coulomb diamonds are well-defined at the Dirac peak. This indicates that a single QD is formed between the source and drain electrodes. The Coulomb diamonds are observed mainly in the p$-$region while an open QD behavior is observed in the n$-$region, showing no Coulomb diamond features. Note that the Dirac peak is at $V_{G}$ $\sim$ $-$2.5 V.

It is rather surprising to observe the quantum confinement effect in the Dirac semimetals because the gapless energy band structure prevents quantum confinement. In conventional semiconductors with finite energy gaps, carriers (either electrons or holes) can be confined by electrostatic barriers with gates. In contrast to massive fermions in conventional semiconductors, Dirac-Weyl fermions can penetrate electrostatic barriers with high transmission probability for normal incident angles (Klein tunneling), making strong confinement difficult.\cite{Katsnelson_Nat_phy2006,Young_Nat_phy2006,Cheianov_PRL2006} However, recent theoretical works propose that Klein tunneling can be suppressed by bending incoming Dirac fermion trajectories with magnetic fields in graphene or Dirac semimetals.\cite{Martino_PRL2007,Kumar_JAP2013,Li_Sci_report_2015} 


\begin{figure}[t]
	\centering
	\includegraphics[width=0.7\textwidth]{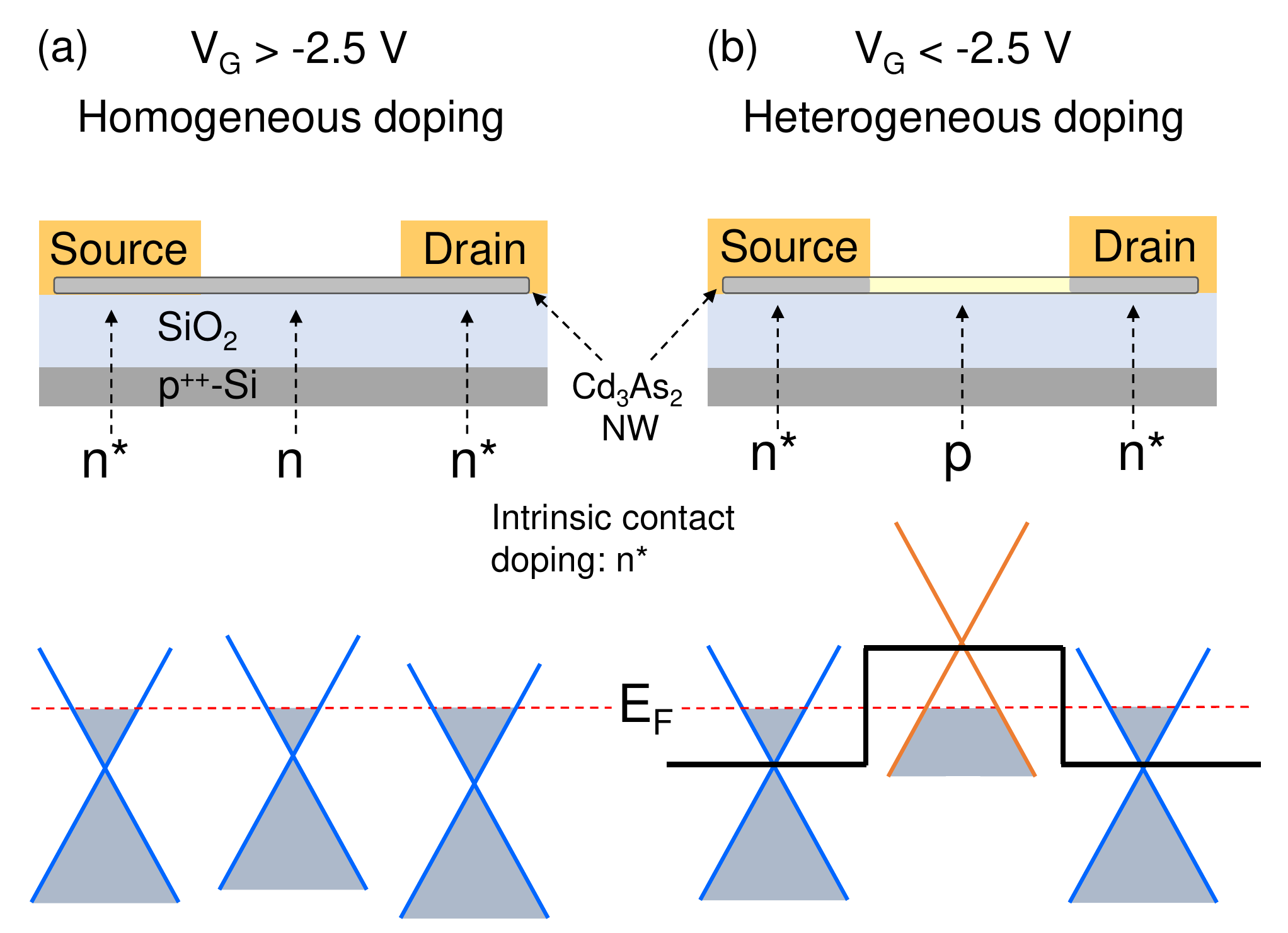}
	\caption{(a) Device schematic with a homogeneous doping in the nanowire by applying $V_{G} \sim -$2.5 V. The entire nanowire is n$-$doped, as mentioned in Figure~2a. Segments of the nanowire under the source$-$drain electrodes (contact doping) are denoted as n$^*$. The lower panel shows the energy band diagram for the n$^*$$-$n$-$n$^*$ configuration, forming an open regime without tunnel barriers. $E_F$ denotes the Fermi energy. (b) Schematic with heterogeneous doping in the nanowire by applying $V_{G} \sim -$2.5 V. The lower panel shows the energy band diagram for the n$^*$$-$p$-$n$^*$ configuration. The QD is p$-$type, while the leads are n$-$type. Quantum confinement is only possible with magnetic fields due to the suppression of the Klein tunneling.}
	
\end{figure}

We here discuss how a single QD can be formed under high magnetic fields in 3D Dirac semimetal nanowires. The differential conductance curve in Figure 2a indicates that the entire nanowire is slightly n$-$doped because the Dirac peak appears at $V_{G} \sim -$2.5 V. As illustrated in Figure~4a, by applying a gate voltage, the carrier density and type of the nanowire between the source$-$drain electrodes can be tuned easily, while segments of the nanowire under the electrodes remain effectively n$-$doped due to less effective gate modulation by the Thomas-Fermi screening.\cite{Weperen_2012,Kammhuber_2017} Hereafter, we denote this effectively n$-$doped nanowire segments by n$^{*}$. When $V_{G}$ $>$ $\sim -$2.5 V (Figure~4a), all segments of the nanowire become n$-$type (n$^*$$-$n$-$n$^*$ configuration), giving rise to an open system without any tunnel barriers. In this regime, the random fluctuating potentials caused by local impurities dominate the phase coherent transports yielding the UCFs as explained. When  $V_{G}$ $<$ $\sim$ $-$2.5 V, the Dirac point of the nanowire segment between the source$-$drain electrodes shifts upward above the Fermi energy $E_{F}$. This allows the nanowire segment to form an n$^{*}$$-$p$-$n$^{*}$ junction, as shown in the schematic in Figure~4b. As mentioned above, this n$^{*}$$-$p$-$n$^{*}$ junction cannot confine Dirac fermions due to Klein tunneling. However, our data shows completely different results, indicating the suppression of Klein tunneling and a strong confinement effect in this regime. Note that the band gap opening by magnetic field as an alternative possibility of quantum dot formations is ruled out because the quantum dots are not formed when the magnetic field is applied along the nanowire axis $[112]$ direction (Supporting Information, Figure S2).

Here we illustrate our scenario for the suppression of Klein tunneling at the n$^{*}$$-$p interface as a function of magnetic field in Figure~5. Figure~5a shows the device schematic and energy diagram of an n$^{*}$$-$p$-$n$^{*}$ junction (lower panel). The yellow area in the lower panel shows an n$^{*}$$-$p interface which is illustrated in Figure~5b. As mentioned above, at $B$ = 0 T, Dirac fermions penetrate the n$^{*}$$-$p interface with high transmission probability due to Klein tunneling (Figure~5b, left panel), which prohibits carrier confinement. However, at weak magnetic fields (middle panel), Dirac fermion trajectories in p and n$^{*}$$-$regimes are bent by the transverse Lorentz force, thus suppressing the transmission probability. Under strong magnetic fields (right panel), the cyclotron motion occurs, further suppressing the Klein tunneling and confining Dirac fermions in the p$-$cavity.    


\begin{figure}[t]
	\centering
	\includegraphics[width=0.8\textwidth]{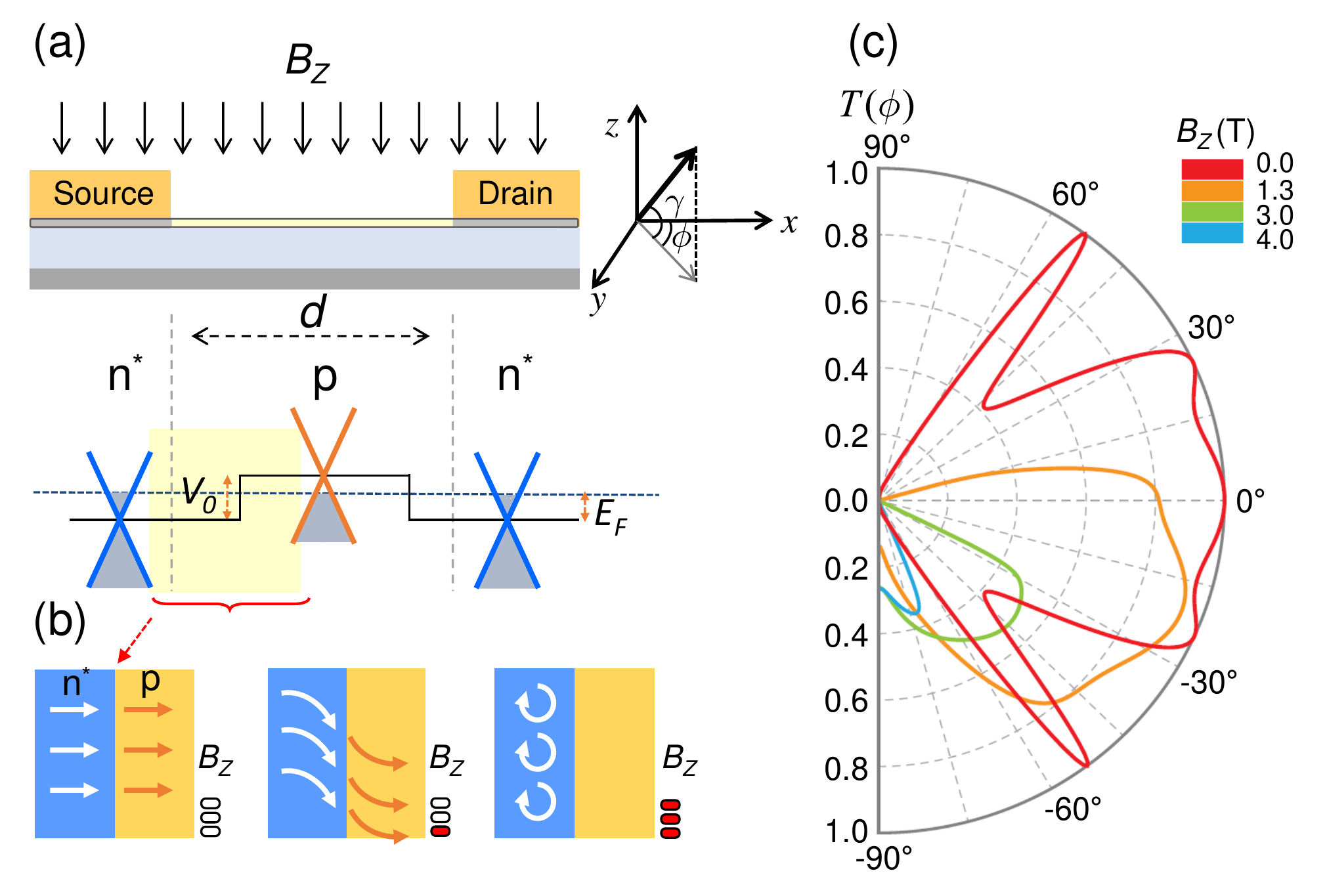}
	\caption{(a) Sketch of device and energy band diagram for n$^{*}$$-$p$-$n$^{*}$ configuration. The yellow box indicates an n$^{*}$$-$p interface (lower panel). $V_0$ is potential height and $E_F$ is the Fermi energy. (b) Illustration of carrier trajectories in the n$^{*}$$-$p interface at different magnetic fields. At $B$ = 0 T (left panel), carriers penetrate the n$^{*}$$-$p junction with high transmission probability. As the magnetic field increases (middle panel), the carrier trajectories are bent in the n$^{*}$ and p$-$ regimes. By applying the magnetic field further (right panel), cyclotron motion occurs, suppressing the Klein tunneling. (c) Transmission probability $T (\phi)$ as a function of carrier incident angle through the p$-$n junction for electrical potential height, $V_0$ = 200 meV. The Fermi energy $E_F$ = 50 meV, Fermi velocity $v_F$ = 2 $\times$ 10$^5$ m/s,\cite{Liu_2015} the central region length $d$ = 120 nm and $\gamma$ = 0.}
\end{figure}

To model this, we then calculate transmission probability $T(\phi)$ of carriers through the p$-$n junction as a function of carrier incident angle for electrical potential height $V_0 = $ 200 meV under magnetic fields. We use the transfer matrix method to calculate the transmission probability by considering three regions shown in the schematic of Figure~5a (See Supporting Information for details). The magnetic field induces a linearly increasing gauge potential along the x-direction. The modulation of the transmission induced by the applied field is due to the decreased overlap between the Fermi surfaces of the source and central segments, which has been discussed in ref.~30. The positions of the Fermi surfaces along the transmission direction is schematically illustrated in Figure~S4. As shown in the Figure~5c, as the magnetic field increases, the angle range that allows transmission shrinks and the electrons become strongly localized. These results support that p$-$n junction in Dirac semimetals under high magnetic fields can behave as a resistive tunnel barrier, confining massless Dirac fermions in either an n$-$p$-$n or p$-$n$-$p junction.\\

\begin{figure}[t]
	\centering
	\includegraphics[width=0.4\textwidth]{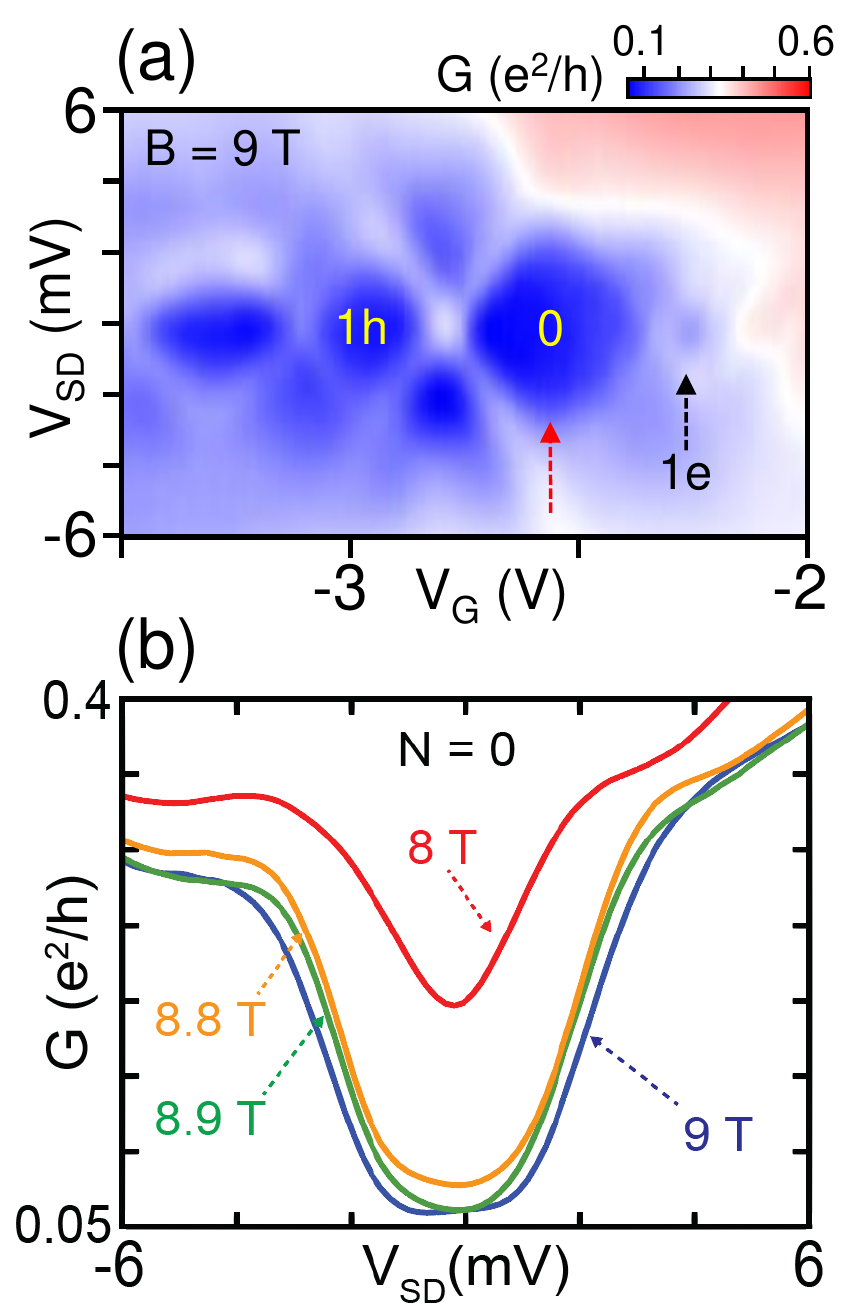}
	\caption{(a) Magnification of the charge stability diagram in Figure~3e taken at $B =$ 9 T. A clean QD behavior is observed in the few electron regime showing  electron filling starting from the empty QD ($N = $ 0) with the first hole (1h) on the left side and the first electron (1e) on the right side. The Coulomb blockade effect is lifted immediately on the electron side, showing no Coulomb diamonds after the first diamond. (b) Conductance measured as a function of $V_{SD}$ at $N = $ 0 (red arrow) for different magnetic fields. As the magnetic field increases, the conductance is strongly suppressed in the Coulomb blockade regime due to the suppression of the Klein tunneling.}
\end{figure}

We next focus on the Coulomb diamonds in detail. Figure 6a shows a magnified plot of Figure 3d taken between $V_{G} = -$2 and $-$3.5 V. The largest Coulomb diamond appears at $V_{G}\sim-2.5$~V, indicating an empty state in the QD (number of carrier, $N = 0$) with the first two adjacent Coulomb diamonds ($N =$ 1h and 1e). The effective energy gap is estimated to be $\sim$ 4 meV from the height of the diamond at $N = $ 0. The QD is sequentially filled with single holes starting from the state $N =$ 1h with one hole in the QD by applying a negative gate voltage. The addition energy required to add a hole into the QD is approximately 2$-$4 meV (Supporting Information, Figure S5). Since Cd$_{3}$As$_{2}$ has a large Lande g-factor of 16,\cite{Li_Nat_comm2015} in a magnetic field, the hole states acquire a large Zeeman energy shift, which causes the energy levels to split and cross each other by $E_{Z} = g\mu B$ ($\sim$ 8.35 meV at $B =$ 9 T).\cite{Hanson_2007} This value is much larger than the Coulomb diamond sizes in Figure~6a so that we cannot observe the Zeeman splitting on the shoulders of the Coulomb diamonds. In addition, the Kondo effect is not observed in our device probably because the Kondo effect is lifted by high magnetic fields. As the temperature increases, the Coulomb oscillations are broadened due to thermal energy (Supporting Information, Figure S6). On the right side of the empty state ($N =$ 0), a small diamond for the first electron ($N =$ 1e) is observed due to an accidental QD formation in the n$^{*}-$n$-$n$^{*}$ configuration. However, the Coulomb blockade in this configuration is immediately lifted as the gate voltage is increased, because the n$-$doped region smoothly merges into the n$^{*}$ lead of the electrodes. In this regime at $B =$ 6 T, the device exhibits a checkerboard pattern on the varying high conductances depending on the gate voltage. This can be interpreted as Fabry-P\'{e}rot interference in a ballistic regime (Supporting Information, Figure S7). Figure~6b shows the conductance measured as a function of $V_{SD}$ at $N = $ 0 (red arrow) for different magnetic fields. As the magnetic field increases, the conductance in the Coulomb blockade regime is strongly suppressed due to the suppression of the Klein tunneling. The result shows that the confinement can further increase under higher magnetic fields.

In summary, we demonstrate single QDs confined with two p$-$n junctions in 3D Dirac semimetal Cd$_{3}$As$_{2}$ nanowires under high magnetic fields. The device can be operated in two different regimes: (i) an n$-$type channel between n$^*$$-$type leads underneath the source-drain contacts, creating an open regime (n$^*$$-$n$-$n$^*$ configuration); (ii) a p$-$type channel in the middle of the nanowire, forming a p$-$type QD (n$^*$$-$p$-$n$^*$ configuration). At zero magnetic field, the quantum confinement effect vanishes in the n$^*$$-$p$-$n$^*$ QD because the Dirac fermions penetrate p$-$n junctions with high transmission probability (Klein tunneling). However, the high magnetic fields bend the Dirac fermion trajectories at the p$-$n junction due to cyclotron motion, preventing the Klein tunneling. This results in a strong confinement at p$-$n junctions of Dirac materials. In this regime, the device shows clean Coulomb diamonds, indicating that a single QD is formed in a Dirac semimetal nanowire. Our experiment allows one to design quantum devices such as QDs or quantum point contacts in 3D Dirac semimetals using magnetic fields.

\begin{acknowledgement}
We thank A. Baumgartner, C. Sch\"{o}nenberger, M.-S. Choi, B.~J. Yang, S.~J. Choi and H.-S. Sim for helpful discussions. This work was supported by the Mid-career Researcher
Program (NRF-2017R1A2B4007862), the Leading Foreign Research Institute Recruitment Program (2012-K1A4A3053565) through the  National Research Foundation of Korea (NRF) grant and DGIST R\&D Program of the Ministry of Science, ICT, Future Planning (17-NT-01 and HRHR2018010137). Research at University of Tokyo was supported by Grant-in-Aid from JSPS
(Nos.16K17481 and 17H01038), MEXT Grant-in-Aid for Scientific Research on Innovative Areas
\textquoteleft Science of hybrid quantum systems\textquoteright (No.15H05868). M.~B.~A.~J. would like to acknowledge the MOE Tier I (NUS Grant No. R-263-000-B98-112), MOE Tier II MOE2013-T2-2-125 (NUS Grant No. R-263-000-B10-112), MOE Tier II MOE2015-T2-1-099 (NUS Grant No. R-380-000-012-112) grants, and NRF-CRP12-2013-01 (NUS Grant No. R-263-000-B30-281) for financial support. N.~N. is supported by JST CREST Grant No. JPMJCR16F1, Japan and by 
JSPS Grant-in-Aid for Scientific Research (No. 24224009, and No. 26103006) from MEXT, Japan, and ImPACT Program of Council for Science, Technology and Innovation (Cabinet office, Government of Japan). X.~-X.~Z. was partially supported by the JSPS Grant-in-Aid for Scientific Research (No. 16J07545). We thank Dr. B.~H. Lee, H.~S. Jang at Center for Core Research Facilities(CCRF) of DGIST for their assistance of device fabrication. 

\end{acknowledgement}

\begin{suppinfo}
	
	Figure S1: TEM and EDS analysis of Cd$_3$As$_2$ nanowires.
	Figure S2: magnetoresistance and temperature dependence of resistance.
	Figure S3: temperature dependence of conductance.
	Figure S4: calculation of transmission probability.
	Figure S5: addition energy of quantum dot.
	Figure S6: temperature dependence of Coulomb stability diagram.
	Figure S7: Fabry-P\'{e}rot like interference at low magnetic field.
		
\end{suppinfo}

\end{document}